# On drug transport after intravenous administration

S.Piekarski (1), M.Rewekant (2)

Institute of Fundamental Technological Research Polish Academy of Sciences (1), Medical University of Warsaw, Poland

## Abstract

A mathematical model of a drug transport after rapid injection is given. It takes into account three processes:

- drug plasma protein binding in central compartment
- transport processes between the central compartment and the peripheral compartment
- elimination of a drug from the central compartment.

.

## Introduction

Drug protein binding is an reversible attachment resulting from electrostatic London-van der Waals's forces, hydrogen bond or their combinations. This reaction satisfy the law of mass action and lasts milliseconds (1). Unbound fraction of drug in vitro is calculated as a quotient of free drug serum concentration and total serum concentration of drug in the closed system (vials space) and given in percents. In design of molecular structures of new drugs, data from drug protein binding in vitro is not often used successfully (2). This approach was questioned by Smith and co-workers, authors of „a free drug hypothesis." They indicated that in human organism (open system), it is free drug concentration, not unbound fraction of drug, that should be considered (3). Concise review of quantification methods used to assessing a degree of drug protein binding was described by Howard and co-workers (4).

Drug molecules weakly bonded with plasma protein may be displaced by strongly bonded molecules. This is an example of protein-drug pharmacokinetic interaction which may be simulated as Christensen and co-workers described (5). Interesting phenomenon of decreasing unbound fraction of drug when the total drug concentration increased in the limit range was recently described by Breshkovsky (6). He suggested the hypothesis of a strong binding site activation by weaker bonding site, present in protein molecules.

Free drug concentration changes in time in a living organism because of dissociation of protein-drug complex, distribution and elimination processes and it has influence on the potency of pharmacological effect.

How drug protein binding changes affect the pharmacokinetic and pharmacodynamic parameters has been the problem discussed in professional literature for many years.

According to some earlier opinions, the drug protein binding in vivo was not clinically significant. There is an opinion that drug protein binding changes affect the clinical outcome due to changes of clearance of a free drug (7). According to Winter there is little clinical evidence of a sharp correlation between free drug concentration and clinical effects (8). Benet and Hoener believed that drug protein binding changes have little effect on bioavailability and clinical outcome of drugs (9).

In 2009 Buur and co-workers demonstrated physiologically based pharmacokinetic model that linked plasma protein binding interactions to drug disposition for SMZ and FLU in a swine. In a study in vivo they confirmed the presence of a drug interaction related to the problem of tissue residue (10). Recently, Schmidt and co-workers made a review of theoretical conceptions related to drug protein binding, describing a status of unbound fraction of drug in basic linear equations of the most important pharmacokinetic parameters such as: clearance, volume of distribution, steady-state concentration of drug and bioavailability and half-life (11). At present, it is necessary to create mathematical models describing not only the dynamic changes of drug protein binding in time but also effects in a range of pharmacokinetic and pharmacodynamic parameters simultaneously (12). There is a need to have a research tool for simulating drug interactions in all effect compartments. In the work presented below we proposed the mathematical modelling of a sequence of processes after intravenous drug administration.

# Basic equations

The standard equations of chemical kinetics, relating free drug concentration $s(t)$, protein concentration $e(t)$ and the drug – protein complex $c(t)$

are

$$\frac{\partial s(t)}{\partial t} = -k_+ e(t)s(t) + k_- c(t) \qquad (1)$$

$$\frac{\partial e(t)}{\partial t} = -k_+ e(t)s(t) + k_- c(t) \qquad (2)$$

$$\frac{\partial c(t)}{\partial t} = k_+ e(t)s(t) - k_- c(t) \qquad (3)$$

where all quantities depend on time and $k_+$, $k_-$ are the reaction rates ($k_+ > 0$, $k_- > 0$).

The above equations describe the time evolution in an uniform system without external sources.

Now, the elimination of a drug is modeled by an additional source term in the balance law (1). In the simplest possible version, such additional source term can by modeled by the term proportional to $s(t)$ and the "strength" of a source can be measured by the proportionality coefficient $\alpha$. Therefore, the resulting equations are now

$$\frac{\partial s(t)}{\partial t} = -k_+ e(t)s(t) + k_- c(t) - \alpha s(t) \qquad (4)$$

$$\frac{\partial e(t)}{\partial t} = -k_+ e(t)s(t) + k_- c(t) \qquad (5)$$

$$\frac{\partial c(t)}{\partial t} = k_+ e(t)s(t) - k_- c(t) \qquad (6)$$

The equations $(4) - (6)$ still do not take into account the transport processes between the central compartment and the peripherial compartment.

Let $V_0$ denote the volume of the central compartment (usually between 5 or 6 litres): then the number of molecules of a free fraction of a drug in a central compartment is $V_0 s(t)$.

Let $N(t)$ denote the nomer of molecules of a drug in a peripherial compartment.

In order to formulate the simplest possible dynamical model of the transport processes between both compartments let us assume that the elimination of a drugi is „switched out".

Then the conservation law for the transport of the free fraction of a drug between both compartments is

$$\frac{\partial}{\partial t}[V_0 s(t) + N(t)] = 0. \qquad (7)$$

Every model of the transport processes between both compartments should identically satisfy the condition (7). The simplest such model depends

On the two parameters and assumes that the volume of the peripherial compartment is $V_{eff}$ and that $N(t)$ molecules are distributed with a uniform density $n(t)$

$$N(t) = V_{eff} n(t) \qquad (8)$$

The explicit form of this model is

$$\frac{\partial}{\partial t} V_0 s(t) = \beta[n(t) - s(t)] \qquad (9)$$

$$\frac{\partial}{\partial t} V_{eff} n(t) = -\beta[n(t) - s(t)] \qquad (10)$$

where the real parameter $\beta$ models the rate of the transport processes between both compartments.

Now one can see that the final system of evolution equations for the drug transport is

$$\frac{\partial s(t)}{\partial t} = -k_+ e(t) s(t) + k_- c(t) - \alpha s(t) + \frac{\beta}{V_0}[n(t) - s(t)] \qquad (11)$$

$$\frac{\partial e(t)}{\partial t} = -k_+ e(t) s(t) + k_- c(t) \qquad (12)$$

$$\frac{\partial c(t)}{\partial t} = k_+ e(t)s(t) - k_- c(t). \tag{13}$$

$$\frac{\partial}{\partial t} V_{eff} n(t) = -\beta [n(t) - s(t)] \tag{14}$$

In future we hope to interpret the solutions of this system in terms of the experimental results.